\documentclass[twocolumn,tighten,times]{aastex631}
\usepackage{amsmath}
\usepackage{amssymb}
\usepackage{xspace,xcolor}
\usepackage{natbib}
\usepackage{outlines}
\usepackage{savesym}
\usepackage{booktabs}
\usepackage{longtable}
\usepackage{tikz}
\usepackage{graphicx}
\usepackage{hyperref}
\savesymbol{tablenum}
\usepackage{siunitx}
\usepackage{comment}
\usepackage{float}
\usepackage{braket}
\usepackage{svg}
\usepackage[normalem]{ulem}
\usepackage{soul}
\usepackage{gensymb}

\newcommand{\tess}{\textit{TESS}}
\newcommand\blfootnote[1]{%
  \begingroup
  \renewcommand\thefootnote{}\footnote{#1}%
  \addtocounter{footnote}{-1}%
  \endgroup
}
\def\one{{\,\sc i}}
\def\two{{\,\sc ii}}
\def\three{{\,\sc iii}}
\def\four{{\,\sc iv}}
\def\five{{\sc v}}
\def\six{{\sc vi}}

\newcommand{\MIT}{Department of Physics and Kavli Institute for Astrophysics and Space Research, Massachusetts Institute of Technology, 77 Massachusetts Avenue, Cambridge, MA 02139, USA}
\newcommand{\WIS}{Department of Particle Physics and Astrophysics, Weizmann Institute of Science, Rehovot 7610001, Israel}

\newcommand{\CWRU}{Department of Astronomy, Case Western Reserve University, 10900 Euclid Avenue, Cleveland, OH 44106, USA}

\newcommand{\JHU}{William H. Miller III Department of Physics \& Astronomy, Johns Hopkins University, 3400 N Charles St, Baltimore, MD 21218, USA}
\newcommand{\STScI}{Space Telescope Science Institute, Baltimore, MD 21218, USA}
\newcommand{\UCSC}{Department of Astronomy and Astrophysics, University of California, Santa Cruz, CA 95064, USA}
\newcommand{\ZJUphy}{Institute for Advanced Study in Physics, Zhejiang University, Hangzhou 310058, China}
\newcommand{\ZJUastro}{Institute for Astronomy, School of Physics, Zhejiang University, Hangzhou 310058, China}
\newcommand{\ARCO}{Astrophysics Research Center of the Open University (ARCO), Department of Natural Sciences, Ra’anana 4353701, Israel}
\newcommand{\CIERA}{Center for Interdisciplinary Exploration and Research in Astrophysics (CIERA), Northwestern University, Evanston, IL 60208,USA}
\newcommand{\NWU}{Department of Physics and Astronomy, Northwestern University, 2145 Sheridan Rd., Evanston, IL 60208, USA}
\newcommand{\Lancaster}{School of Physics and Astronomy, Lancaster University, Lancaster, LA1 4YB, UK}
\newcommand{\FCLA}{French-Chilean Laboratory for Astronomy, IRL 3386, CNRS and Instituto de Astrofísica, Pontificia Universidad Católica de Chile, Casilla 306, Santiago, Chile.}
\newcommand{\IAPCNRS}{Institut d'Astrophysique de Paris, CNRS-Sorbonne Universit\'e, 98 bis boulevard Arago, F-75014 Paris, France}
\newcommand{\UCAS}{School of Astronomy and Space Science, University of Chinese Academy of Sciences, Beijing 100049, People’s Republic of China}
\newcommand{\NAOCAS}{National Astronomical Observatories, Chinese Academy of Sciences, Beijing 100101, People’s Republic of China}
\newcommand{\umel}{School of Physics, The University of Melbourne, VIC 3010, Australia}
\newcommand{\dark}{DARK, Niels Bohr Institute, University of Copenhagen, Jagtvej 128, 2200 Copenhagen, Denmark}
\newcommand{\uhawaii}{Institute for Astronomy, University of Hawaii, 2680 Woodlawn Drive, Honolulu, HI 96822, USA}
\newcommand{\cfa}{Center for Astrophysics | Harvard \& Smithsonian, 60 Garden Street, Cambridge, MA 02138, USA}
\newcommand{\uiuc}{Department of Astronomy, University of Illinois at Urbana-Champaign, 1002 W. Green St., IL 61801, USA}
\newcommand{\cas}{Center for Astrophysical Surveys, National Center for Supercomputing Applications, Urbana, IL, 61801, USA}
\newcommand{\DARK}{DARK, Niels Bohr Institute, University of Copenhagen, Jagtvej 128, 2200 Copenhagen, Denmark}
\newcommand{\INAF}{INAF, Osservatorio Astronomico di Capodimonte, Salita Moiariello 16, I-80131 Naples, Italy}

\begin{document}

\title{Constraints on the Progenitor of SN~2023bee from its Nebular-Phase Spectra}

\author[0000-0002-8436-5431]{Sophie von Coelln}\blfootnote{Corresponding author: Sophie von Coelln\\ \href{mailto:sophie.von.coelln@gmail.com}{sophie.von.coelln@gmail.com}}
\affiliation{\CWRU}
\author[0000-0001-5233-6989]{Qinan~Wang}%
\affiliation{\MIT}
\affiliation{\WIS}
\author[0000-0003-2445-3891]{Matthew R. Siebert}
\affiliation{\STScI}
\author[0000-0003-0853-6427]{Ping Chen}
\affiliation{\ZJUphy}
\affiliation{\ZJUastro}
\author[0000-0002-4410-5387]{Armin Rest}
\affiliation{\STScI}
\affiliation{\JHU}
\author{Luc Dessart}
\affiliation{\FCLA}
\affiliation{\IAPCNRS}
\author[0000-0002-0632-8897]{Yossef Zenati}
\affiliation{\JHU}
\affiliation{\ARCO}
\author[0000-0001-9494-179X]{Georgios Dimitriadis}
\affiliation{\Lancaster}
\author[0000-0002-4449-9152]{Katie~Auchettl}
\affiliation{\umel}
\author[0009-0007-7564-4408]{Siyuan Chen}
\affiliation{\cfa}
\author[0000-0002-5680-4660]{Kyle~W.~Davis}
\affiliation{\UCSC}
\author[0000-0001-5486-2747]{Thomas~de~Boer}
\affiliation{\uhawaii}
\author[0000-0002-2445-5275]{Ryan~J.~Foley}
\affiliation{\UCSC}
\author[0000-0002-3653-5598]{Avishay Gal-Yam}
\affiliation{\WIS}
\author[0000-0003-3953-9532]{Willem~B.~Hoogendam}
\affiliation{\uhawaii}
\author[0000-0003-1059-9603]{Mark~E.~Huber}
\affiliation{\uhawaii}
\author[0000-0001-9695-8472]{Luca~Izzo}
\affiliation{\DARK}
\affiliation{\INAF}
\author[0000-0002-5740-7747]{Charles~D.~Kilpatrick}
\affiliation{\CIERA}
\affiliation{\NWU}
\author[0000-0002-7272-5129]{Chien-Cheng~Lin}
\affiliation{\uhawaii}
\author[0000-0002-7866-4531]{Chang Liu}
\affiliation{\CIERA}
\affiliation{\NWU}
\author[0000-0002-9438-3617]{Thomas~Bernard~Lowe}
\affiliation{\uhawaii}
\author[0000-0002-2260-4628]{Gautham Narayan}
\affiliation{\uiuc}
\affiliation{\cas}
\author[0000-0002-7559-315X]{César~Rojas-Bravo}
\affiliation{\UCAS}
\affiliation{\NAOCAS}
\author[0000-0002-9158-750X]{Aidan~Sedgewick}
\affiliation{\dark}
\author[0000-0002-2798-2943]{Koji~Shukawa}
\affiliation{\JHU}
\author[0000-0002-1341-0952]{Richard~J.~Wainscoat}
\affiliation{\uhawaii}

\begin{abstract}

We present late-time photometry ($\sim$100--400~days after peak) and nebular-phase optical/NIR spectroscopy ($\sim$100--360~days) of the Type Ia supernova (SN~Ia) SN~2023bee. 
This SN~Ia was observed by \tess\ immediately before and after explosion and showed a relatively weak early excess light-curve feature detected in multiple UV and optical bands. Possible explanations for this early excess feature include interaction between the SN ejecta and a non-degenerate companion, or the decay of radioactive isotopes synthesized near the ejecta surface in a sub-$\text{M}_\text{Ch}$ double-detonation scenario. However, the degeneracies and incompleteness of existing models have so far prevented a definitive conclusion about its progenitor system. 
Nebular-phase spectra provide a complementary probe of SN~Ia progenitors by constraining stripped H/He-rich material and iron-group abundances sensitive to the progenitor white dwarf density. 
We report the non-detection of narrow H/He features in nebular phase optical/NIR spectra of SN~2023bee, placing upper limits of $m^{\textrm{stripped}}_\textrm{H}\lesssim5.0\times10^{-3}\textit{M}_\odot$ and $m_\textrm{He}^{\textrm{stripped}}\lesssim1.2\times10^{-4}\textit{M}_\odot$. These limits strongly disfavor a non-degenerate companion and suggest that companion interaction is unlikely to explain the early excess of SN~2023bee. 
We also performed line-diagnostic measurements on the $7200$~\AA\ complex using multi-Gaussian fitting and continuum estimation, with the inferred velocities validated against NIR spectrum at the similar phase. The resulting low Ni-to-Fe mass ratio, $m_\textrm{Ni}/m_\textrm{Fe}\lesssim 0.06$, is broadly consistent with a sub-$M_{\text{Ch}}$ double-detonation progenitor system.

\end{abstract}


\section{Introduction}\label{sec:intro}
Type Ia supernovae (SNe~Ia) are generally believed to originate from the thermonuclear explosion of a white dwarf (WD) star, yet the nature of their progenitor systems remains uncertain. In particular, the properties of the progenitor system specifically the type of companion star, as well as the physical conditions leading to explosion, are not yet fully established. 
This uncertainty is particularly important given the use of SNe~Ia as standardized candles for measuring the expansion history of the Universe \citep{Riess1998, Perlmutter1999}.
Different progenitor and explosion scenarios are expected to produce different early-time signatures, including excess emission shortly after explosion, but uncertainties in theoretical modeling and limitations in early-time follow-ups often make it hard to exclusively connect them with certain progenitor systems. Nebular-phase spectroscopy provides complementary constraints on the inner ejecta and can help distinguish among otherwise degenerate scenarios.

The complex physics behind SN Ia explosions has been investigated for decades \citep[e.g.,][]{Hoyle_Fowler1960, Livio_Mazzali18, Ruiter2020, ZhangWei_Ropke_Han23, RuiterSeitenzahl24}. The two main channels for SN Ia progenitor systems are single degenerate (SD) and double degenerate (DD) systems. In the SD scenario, the WD explodes due to a thermonuclear runaway triggered as the WD accretes mass from a nondegenerate companion star near the Chandrasekhar mass \citep[$\text{M}_\text{Ch}$;][]{WhelanIben1973}. In the DD scenario, mass transfer or a merger event between two WDs results in a thermonuclear runaway, leading to a standard or peculiar supernova explosion \citep{IbenTutukov1984, Fink+10,Hillebrandt+13,RuiterSeitenzahl24}. These broad progenitor classes encompass a range of binary configurations and evolutionary pathways that may produce both normal and peculiar SNe~Ia.

A separate but related question concerns the mass and ignition conditions of the exploding WD. Near-Chandrasekhar-mass ($M_{\rm Ch}$) explosions occur when the WD grows to approximately $M_{\rm Ch}$ before thermonuclear runaway \citep[e.g.,][]{Seitenzahl13, Casabona_Fisher24}, whereas sub-$M_{\rm Ch}$ explosions occur at substantially lower masses. Sub-$M_{\rm Ch}$ explosions could be triggered by double detonation \citep[e.g.][]{Blondin17, Shen18,Polin2019,Perets_Zenati+19, Boos2021} or a merger \citep[]{Pakmor12, Pakmor+13, Ruiter13}.
Other proposed channels include the core-degenerate (CD) scenario \citep{Iben_Livio93, Sandquist+98, Kashi_Soker11, WangB+17},  distributed burning realized in turbulently driven detonation regime \citep{WoosleyS07, PoludnenkoA+11, BrookerE+21, Zenati_Fisher23}, 
and all versions of collisions such as head-on or impact parameter between double WDs \citep{Bens_Hills_Thielemann1989, Raskin+2010, Kushnir13, DongS+15, Papish_Perets16, GlanH+25}. More generally, the thermonuclear burning may proceed through deflagration, detonation, or transitions between burning regimes, further broadening the range of possible SN~Ia explosion outcomes.

One potential way to distinguish between SD and DD scenarios is the early light curve: we expect the early light curve of SNe~Ia to rise roughly following a power law, with the power-law index close to $2$ in the absence of additional emission \citep{Arnet1982, Hoogendam24}. Several mechanisms can produce an early excess, or ``bump'' in extreme cases, including ejecta--companion interaction in the SD scenario \citep{Kasen2010}; shallow or asymmetric $^{56}$Ni distributions \citep{PiroNakar2013, Magee2020, Magee_Maguire20}; circumstellar interaction \citep{PiroMorozova2016, Moriya2023}; and radioactive products of He-shell detonations in double-detonation models \citep{Polin19}. Distinguishing among these possibilities requires densely sampled observations within the first hours to days after explosion. Such data are difficult to obtain because most SNe are discovered after the earliest phases, while rapid follow-up is not always available. In addition, uncertainties in explosion and radiative-transfer models, together with degeneracies among model parameters and viewing angle, limit the extent to which even well-observed early excesses can be uniquely interpreted.

In addition to the observational features like the early light curve, nebular phase spectroscopy provides another window into the progenitor nature of SNe~Ia. In an SD scenario, where the SN ejecta shock the nondegenerate companion, they will also strip material from the surface of the companion, which will be evident in spectra taken during the nebular phase of the SN \citep{Marietta00, Leonard07, Boehner18, Botyanszki17}. The stripped material will produce strong and narrow emission lines on top of an otherwise standard SN Ia spectrum. In particular, we expect to see some or all of H$\alpha$, H$\beta$, He I $\lambda$5875, and He I $\lambda$6678, varying based on the type of companion star and the characteristics of the binary system. 

Moreover, the density attained during thermonuclear burning affects the degree of electron capture and therefore the production of stable neutron-rich iron-group isotopes, providing a potential diagnostic of the SN Ia explosion mechanism \citep{hoflich04, Maguire18}. In particular, nebular [Fe,\two] and [Ni,\two] emission can be used to constrain the stable Ni-to-Fe abundance ratio, which is expected to differ between near-$M_{\rm Ch}$ and sub-$M_{\rm Ch}$ explosion scenarios \citep{flors20, Siebert23}. We discuss these diagnostics and their associated uncertainties in detail in Section~\ref{sec:7300A}.

A number of SNe~Ia have been found with early flux excess, including SNe 2017cbv, 2018oh, 2021aefx  \citep[][]{HosseinzadehG+17_2017cbv,dimitriadis2018k2,Li2018oh, shappee2019seeing, Ashall2022}. These SNe range from normal to overluminous, and their flux excess is normally blue and bright in UV with a few red exceptions \citep[such as SN~2018aoz;][]{2023ApJ...946....7N}. The shape of 2018oh's early light curve can be fit by the SD companion shocking model, the shallow $^{56}$Ni model, or the sub-$\text{M}_\text{Ch}$ double detonation model, but none of these models can explain both the blue early excess and the lack of narrow emission lines in nebular spectroscopy. SNe 2017cbv and 2021aefx have early UV luminosity measurements from \textit{Swift}, which are too low to align with the bright UV bump expected from the SD companion-shocking model. In addition, SD models have been largely excluded for SNe 2017cbv and 2018oh by multiple studies, due to insufficient hydrogen and helium lines in their nebular spectra \citep{Sand2018, Dimitriadis2019, Tucker19}. Alternative explanations also face observational challenges, models containing an outer $^{56}$Ni shell can reproduce the early optical excess but fail to match the color and spectral evolution, predicting excessive line blanketing and an unobserved absorption feature around $3700-4000$\AA\ \citep{Magee_Maguire20}. Interaction with H/He-poor CSM remains a possible explanation for SN~2017cbv, and interaction with disk-originated matter has been shown to reproduce the early rise of SN~2018oh \citep[e,g,.][]{HosseinzadehG+17_2017cbv}. None of the current models are sufficient to describe potential progenitor systems for all SNe~Ia with early excess flux.

In this paper, we examine the late-time spectra of SN 2023bee, a normal SN Ia observed by the Transiting Exoplanet Survey Satellite (\textit{TESS}) discovered with a weak early excess. It was discovered 2 days post-explosion and also observed photometrically for 12 days pre-explosion with a 10-minute cadence, identified with 
. This extensive observation allows us to closely investigate the early flux and nebular spectrum of SN 2023bee and their effects on the progenitor system model for the SN. \cite{Hosseinzadeh2023} presents an early spectral time series and radio observations of the SN, concluding that the companion-interaction model in \cite{Kasen2010} best aligns with the observed early excess flux, and noting that, under such a model, the ejecta should strip material from the companion, which could be observed in nebular-phase spectroscopy. \cite{Wang23} presents a more detailed examination of the photometric data, including the \tess\ photometry, finding the limitations of existing models in explaining the early excess of SN~2023bee across all UV-optical bands. This motivates additional late-time observations to better constrain the progenitor system of SN~2023bee. 
In \S~\ref{sec:data} we present the acquisition, reduction and calibration of our photometric and spectroscopic data. In \S~\ref{sec:H/He} we analyze the H/He features in the optical and NIR spectra, and place upper limits on the H/He-rich material stripped from a companion. In \S~\ref{sec:7300A} we estimate the iron-group abundance ratio to constrain the central density and the progenitor system. We present the implications for the progenitor system in \S~\ref{sec:discussion}. All the data presented in this paper will be made public via WISeREP \footnote{https://www.wiserep.org/object/22562}\citep{wiserep}.

\begin{figure*}
    \centering
    \includegraphics[width=0.9\linewidth]{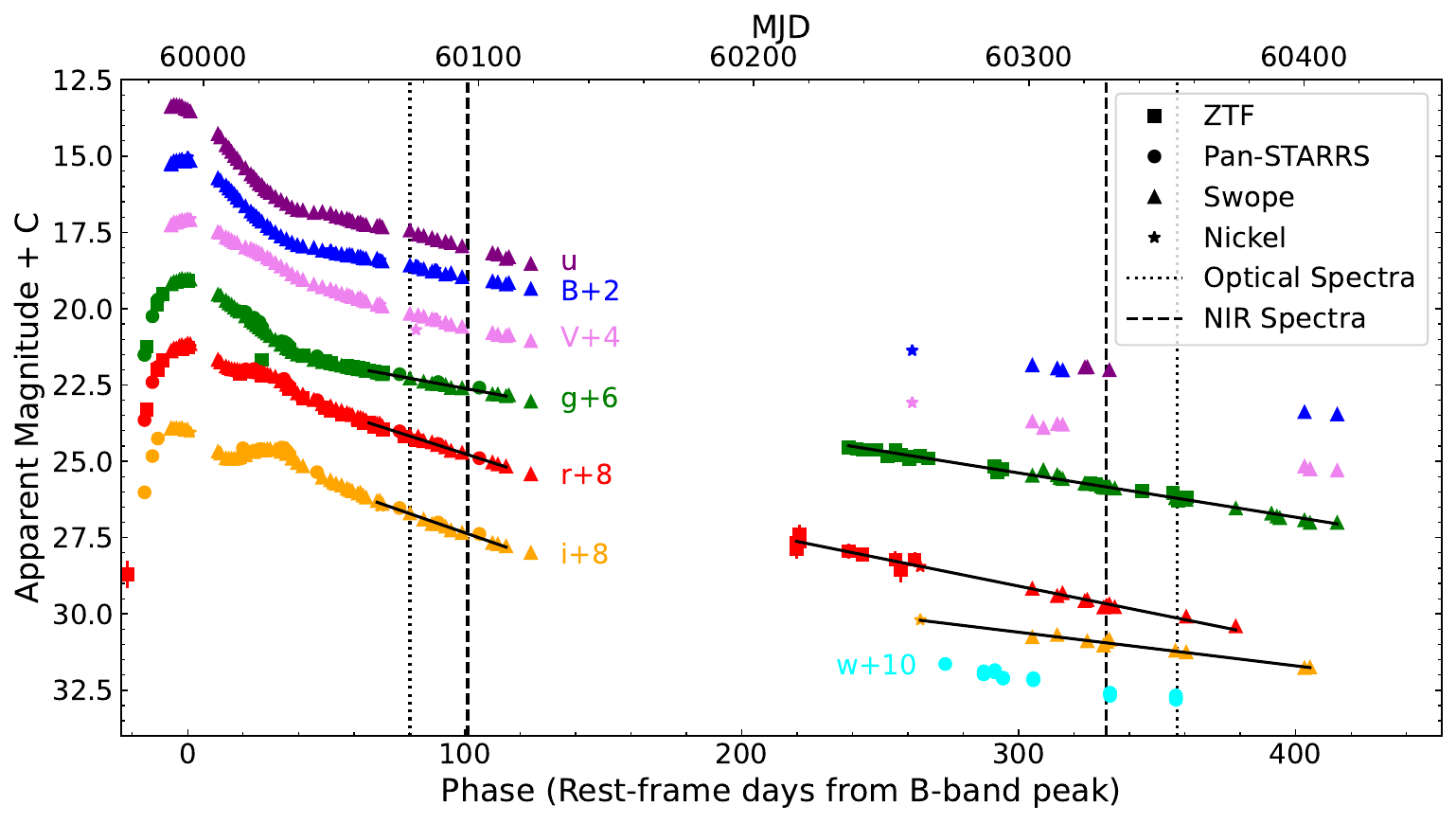}
    \caption{Multi-band light curves of SN~2023bee offset for display. The filter names and corresponding offsets are labeled in the break in the data. 
    We use the interpolated photometry in $gri$-bands to flux calibrate the spectrum. The linear interpolation of the photometry are shown as solid lines between two MJD ranges covering the spectra taken, i.e., MJD 60060-60110 and 60200-60425. The solid and dashed black lines indicate the days the optical and NIR spectra were taken respectively.}
    \label{fig:interpolation}
\end{figure*}

\section{Data Acquisition, Reduction and Calibration}\label{sec:data}

SN 2023bee was discovered at coordinates $\alpha=08$:56:11.63 and $\delta=-03\degree19'32.06''$ (J2000.0). It peaked at 59992.58 MJD with an apparent magnitude of $m_B=13.041\pm0.001$ mag, and absolute magnitude of $M_B =-19.6\pm0.2$ mag \citep{Wang23}. It is located in galaxy NGC 2708, an intermediate spiral galaxy at $z=0.0067\pm0.0005$, and is 93.97" from the center of the galaxy. It has a distance modulus $\mu=32.5\pm0.2$ mag \citep{Pisano2002}. We expect negligible host extinction given SN~2023bee's large projected offset from host nucleus. In this paper, we use $E(B-V)_{MW}=0.145$ for Milky Way extinction based on the map given in \citet{Schlafly2011}, and dust extinction law from \citet{Fitzpatrick99}.   

\subsection{Late-time Photometry}


We obtain the late-time photometry of SN~2023bee with the Nickel 1.0-m telescope at the Lick Observatory in $BVri$ bands, the Swope 1.0-m telescope at Las Campanas Observatory in $uBVgri$ bands, and
the Pan-STARRS1 1.8~m telescope at Haleakala on Maui, Hawai'i in $w$ bands \citep{2016arXiv161205560C}, as part of the Young Supernova Experiment (YSE; \cite{jones21}). We also include the publicly available data from the Zwicky Transient Facility \citep[ZTF;][]{2019Bellm_ZTF,2020Dekany_ZTFObs} in $gr$ bands. 
Standard reductions for the DECam and PS1 images are performed by the NOIRLab community pipeline \citep{valdes_decam_2014} and the PS1 Image Processing Pipeline (IPP), respectively \citep{2020ApJS..251....3M,2020ApJS..251....5M,2020ApJS..251....6M,2020ApJS..251....4W}. 
These images are then taken as input to the \texttt{photpipe} pipeline \citep{rest_testing_2005,rest_cosmological_2013}, which re-determines the zero points by comparing DoPHOT PSF photometry from each image to the Pan-STARRS DR1 catalog \citep{PS1catalogue}, convolves and subtracts a template image from the survey image, and performs forced photometry on the resulting difference images to create transient light curves.
As described in \cite{kilpatrick2018}, all Swope image processing and optical photometry was performed using {\tt photpipe} \citep{rest_testing_2005}, flat-fielding, image stitching, and photometric calibration. $BgVri$ photometry were calibrated using standard sources from the Pan-STARRS DR1 catalog \citep{PS1catalogue}, while the u-band data were calibrated using SkyMapper $u$-band standards \citep{onken2019}, transformed into the Swope natural system \citep{Krisciunas} with the Supercal method \citep{Scolnic_supercal}. Nickel $BVri$ imaging was reduced with \texttt{photpipe} \citep{rest_testing_2005,kilpatrick2018} following the procedures described for Nickel and LCO data in \citet{Gagliano2022} and \citet{Kilpatrick2023}. We applied bias and flat-field frames obtained on the same night and in the same instrumental configuration, aligned each frame to 2MASS astrometric standards \citep{Skrutskie2006}, and regridded the images with SWarp \citep{Bertin2010} to remove geometric distortion. PSF photometry was performed with a custom version of DoPHOT \citep{Schechter1993}. The $gri$ photometry was calibrated to Pan-STARRS DR1 standards \citep{PS1catalogue} in the same field, and Johnson $BV$ magnitudes were obtained by transforming those PS1 standards following \citet{Tonry2012}.

\begin{deluxetable}{cccccc}
\tablecaption{Photometric measurements of SN 2023bee. The complete table is available in machine-readable form.\label{tab:photometry}}
\tablehead{
\colhead{MJD} & \colhead{Phase\tablenotemark{a}} & \colhead{Filter} &
\colhead{$m$} & \colhead{$\sigma_m$} & \colhead{Telescope/Instrument} \\
\colhead{(day)} & \colhead{(day)} & \colhead{} & \colhead{(mag)} & \colhead{(mag)} & \colhead{}
}
\startdata
59976.998 & -15.478 & o & 16.793 & 0.030 & ATLAS/HKO \\
59977.018 & -15.458 & o & 16.764 & 0.017 & ATLAS/HKO \\
59978.552 & -13.935 & g & 15.510 & 0.010 & Pan-STARRS1/GPC1 \\
59978.552 & -13.935 & r & 15.649 & 0.011 & Pan-STARRS1/GPC1 \\
59978.553 & -13.934 & i & 16.017 & 0.014 & Pan-STARRS1/GPC1 \\
59978.554 & -13.933 & z & 15.848 & 0.014 & Pan-STARRS1/GPC1 \\
59978.555 & -13.932 & y & 16.105 & 0.034 & Pan-STARRS1/GPC1 \\
59979.294 & -13.198 & g & 15.242 & 0.024 & P48/ZTF-Cam \\
59979.401 & -13.091 & r & 15.309 & 0.025 & P48/ZTF-Cam \\
59981.430 & -11.076 & g & 14.250 & 0.005 & Pan-STARRS1/GPC1 \\
59981.431 & -11.075 & r & 14.406 & 0.005 & Pan-STARRS1/GPC1 \\
59981.431 & -11.075 & i & 14.829 & 0.005 & Pan-STARRS1/GPC1 \\
59981.432 & -11.074 & z & 14.812 & 0.007 & Pan-STARRS1/GPC1 \\
59981.433 & -11.073 & y & 15.055 & 0.017 & Pan-STARRS1/GPC1 \\
59983.204 & -9.314 & g & 13.868 & 0.016 & P48/ZTF-Cam \\
59983.337 & -9.181 & r & 14.000 & 0.015 & P48/ZTF-Cam \\
59983.397 & -9.122 & g & 13.704 & 0.002 & Pan-STARRS1/GPC1 \\
59983.398 & -9.121 & r & 13.907 & 0.004 & Pan-STARRS1/GPC1 \\
59983.398 & -9.121 & i & 14.254 & 0.005 & Pan-STARRS1/GPC1 \\
59983.399 & -9.120 & z & 14.360 & 0.008 & Pan-STARRS1/GPC1 \\
\enddata
\tablenotetext{a}{Rest-frame phase relative to the adopted B-band maximum at $\mathrm{MJD}_{\rm peak}=59992.58$.}
\end{deluxetable}

\subsection{Optical and NIR Spectroscopy}

We report two collected late-time optical spectra of SN~2023bee in this paper. One optical spectrum was taken by the Kast spectrograph on the Lick 3\,m (Shane) telescope on 2023 May 11 (MJD 60075), $\sim+82$ days after the peak in rest frame. Another optical spectrum was taken by Keck telescope with LRIS spectrograph on Feb 14 2024 (MJD 60354), $\sim+359$ days after the peak in nebular phase.
To reduce the Kast and LRIS data, we used the {\tt UCSC Spectral Pipeline}\footnote{\url{https://github.com/msiebert1/UCSC\_spectral\_pipeline}} \citep{Siebert20}, a custom data-reduction pipeline based on procedures outlined by \citet{Foley03}, \citet{Silverman2012}, and references therein.  The two-dimensional (2D) spectra were bias-corrected, flat-field corrected, adjusted for varying gains across different chips and amplifiers, and trimmed.  One-dimensional spectra were extracted using the optimal algorithm \citep{Horne86}.  The spectra were wavelength-calibrated using internal comparison-lamp spectra with linear shifts applied by cross-correlating the observed night-sky lines in each spectrum to a master night-sky spectrum.  Flux calibration and telluric correction were performed using standard stars at a similar airmass to that of the science exposures.  We combine the red and blue sides by scaling one spectrum to match the flux of the other in the overlap region and use their error spectra to correctly weight the spectra when combining.  More details of this process are discussed elsewhere \citep{Foley03, Silverman2012, Siebert20}.
Two NIR spectra were taken by Magellan telescope with the folded port infrared echellette \citep[FIRE; ][]{MagellanFIRE}. They were reduced and extracted with \texttt{PypeIt} \citep{pypeit:joss_pub, pypeit:zenodo}. We chose to rebin the LRIS and FIRE spectrum to $3.5$\AA\ and $1.67$\AA\ respectively to approximately match their actual spectroscopic resolution power. The Kast spectrum was not rebinned as it only serve as photometric calibrator in later analysis.

\begin{figure*}
    \centering
    \includegraphics[width=0.95\linewidth]{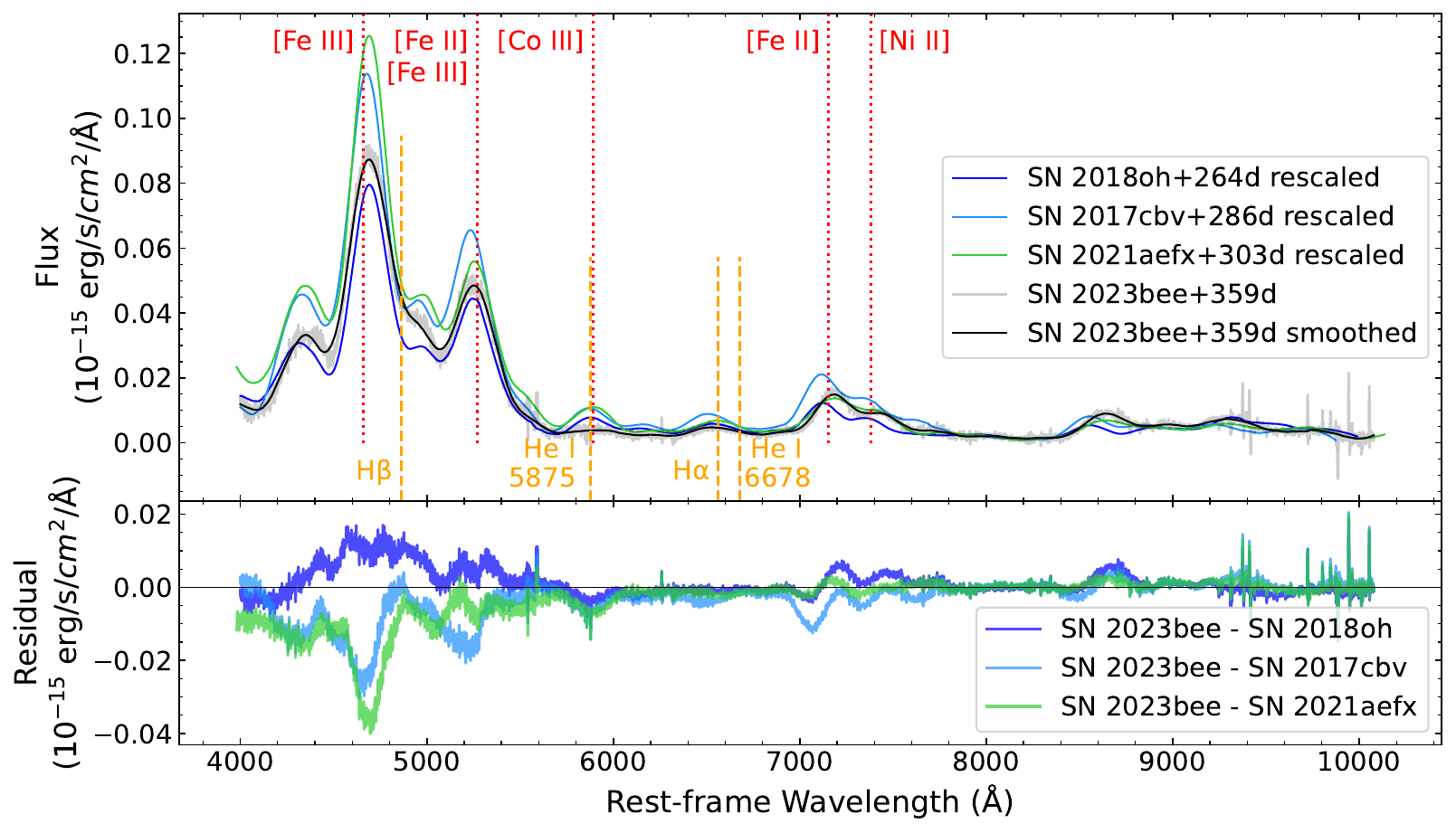}
    \caption{LRIS spectrum of SN 2023bee taken on +359 days, shown in gray and shown smoothed in black, compared with smoothed and normalized SN 2017cbv +286 days in light blue, smoothed SN 2018oh +264 days in dark blue, and SN 2021aefx in green. Those spectra were multiplied by 0.093, 1.445, and 0.177, respectively, for display only to rescale them to match with the smoothed SN~2023bee spectrum between 7500 and 8500\AA. The SN~2021aefx data is B-band data bluewards of 6150\AA, r-band data redwards of 6350\AA, and averaged B- and r-band data within that wavelength range. Only data for wavelengths above 4000\AA~are shown to see features in the remaining spectrum more clearly than the noisy low wavelengths of SN 2023bee allow. The phases are relative to the B-band maximum. We label the rest wavelength of H/He features in yellow, and the detected iron-group features in red. } 
    \label{fig:comparison}
\end{figure*}

The spectrum at +359 day for SN 2023bee is shown in light gray in Figure~\ref{fig:comparison}; overplotted in black is the smoothed spectrum, and absolute flux calibration is discussed in Section~\ref{sec:specfluccal}. Also shown in blue are the normalized, redshift-corrected, and similarly smoothed nebular phase spectra of three SNe~Ia with early excess, i.e. SN~2017cbv, SN~2018oh and SN~2021aefx, at +286 , +264 and +303 days respectively. The LRIS optical spectrum taken on +359 days shows blends of broad features from ionized Fe, Co and Ni, including the strong [Fe\,\three] emission around 4700\AA\ and [Co\,\three] around 5900\AA. Another noticeable feature is the blended region around 7200\AA\ composed of [Fe\,\two] and [Ni\,\two] multiplets, and potentially existing [Co\,\three] and [Ca\,\two] features. As shown in Figure~\ref{fig:comparison}, SN~2023bee show no detectable narrow H/He features that are expected in the SD scenario, consistent with other well-studied SNe~Ia exhibiting early excesses, as illustrated in the bottom residual panel. 

Similar to optical spectra, the +101 day NIR spectrum of SN~2023bee shows strong broad iron-group line features including [Co\,\two] 10190\AA, [Fe\,\two] 12570\AA\ and [Co\,\two] 15470\AA. At +333 day, the iron-group line features become more prominent relative to weaker continuum, with broad [Fe\,\two] 16440\AA\ feature also appearing.
On the other hand, no significant H/He feature can be seen on these nebular phase spectra of SN~2023bee.
The analysis on the line features around 7200\AA\ and in NIR will be discussed in Section~\ref{sec:7300A}.

\subsection{Spectroscopic flux calibration}\label{sec:specfluccal}

To obtain an accurate line intensity measurement for further physical analysis, the spectra need to be flux calibrated. The spectroscopic synthetic magnitude in each filter was determined with the \textsc{pyphot} package from the fluxes and wavelengths of the two optical spectra. The photometric magnitude was then found via linearly interpolating late-time photometry from the Swope $gri$ filters to infer the accurate magnitudes of SN~2023bee in each band on the date the spectrum was taken, as shown in the Figure~\ref{fig:interpolation}. 
We report the interpolated magnitudes to be $m=20.21, 22.14$ and $21.23$ mag in the $gri$ bands respectively at +359 day, in good agreement with neighboring Swope photometric points with $m_g = 20.208$ mag and $m_i = 21.2$ mag at +358 day, $m_r = 22.089$ mag at +362 day.
Computing the difference between the photometric and spectroscopic synthetic magnitude in each band yields -37.306, -37.231, and -37.188 mags in $g$, $r$, and $i$ bands, respectively. The $\sim0.1$ mag difference between $g$ and $r/i$-band offsets may arise from a relative flux-calibration mismatch between the blue and red arms of LRIS, and we therefore include this difference as a component of the photometric calibration uncertainty. 
From there, we normalize the spectra with the average difference between the photometric and synthetic magnitude in $gri$-bands, and then correct for the distance modulus. Finally, the rest-frame spectrum was smoothed with a third-order Savitzky-Golay filter with window size of 100\AA. The small window size means this smoothed out noise while maintaining more significant features, allowing us to better identify any noticeable emission lines that may come from the remaining H and He from the progenitor. 
We estimate that the flux conversion scale of the +359 day optical spectrum has a $\sim4.7\%$ uncertainty associated with it given the uncertainties and scatter in photometric measurements.

\begin{figure*}
    \centering
    \includegraphics[width=0.95\linewidth]{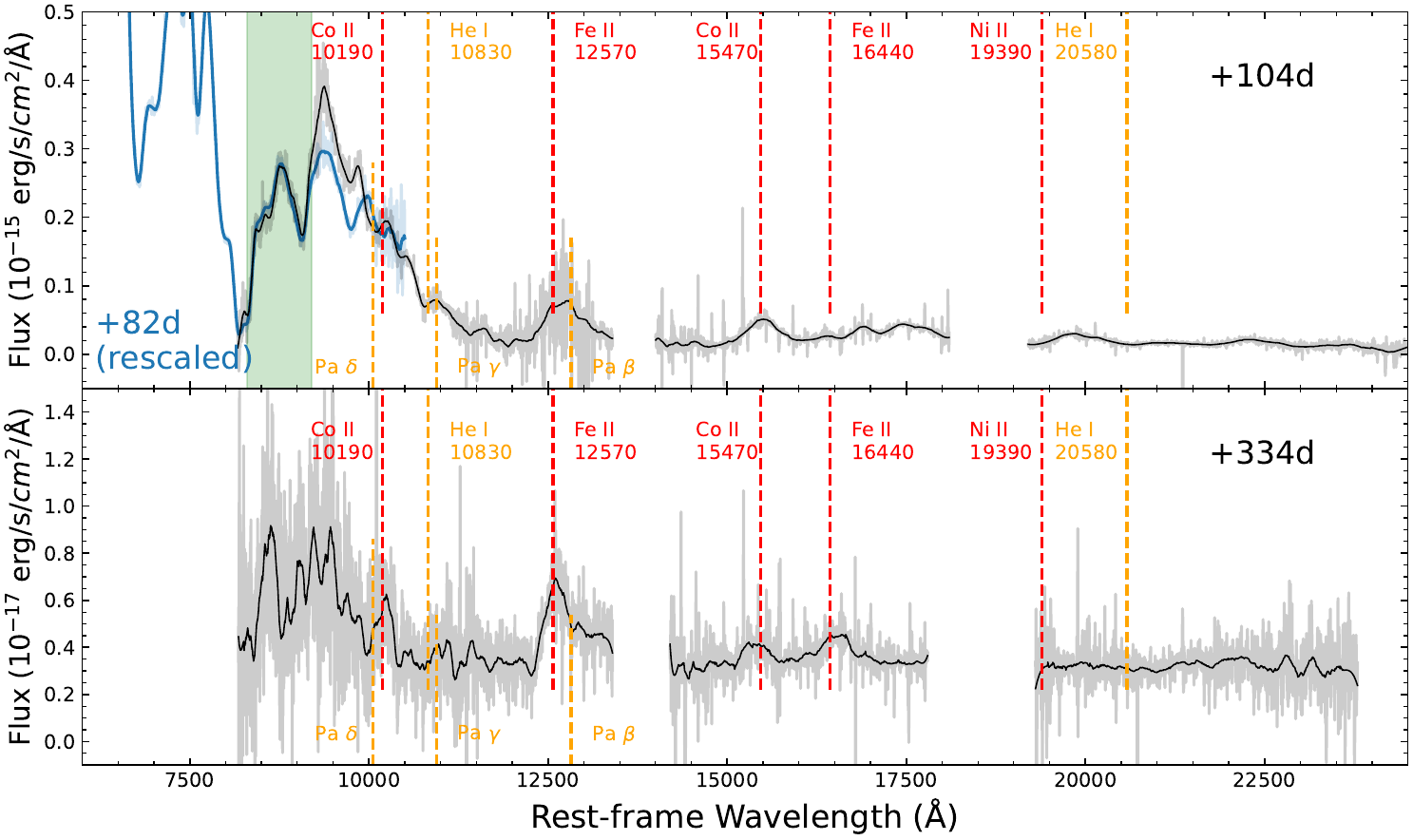}
    \caption{FIRE spectrum of SN 2023bee taken on +103.7 day (top) and +334.3 day (bottom). The Kast optical spectrum obtained at +82.0 days, used for photometric calibration, is also shown after being scaled according to the evolution of the $i$-band magnitude between the two epochs. The shaded regions indicate the wavelength intervals used for photometric calibration that are relatively free of telluric contamination. All spectra have been corrected for redshift and extinction.} 
    \label{fig:NIRcomparison}
\end{figure*}

Afterwards, we first flux-calibrate the optical spectrum on +80 day with $gri$ photometry in a similar way, and then flux-calibrate the adjacent +101 day NIR spectra based on the overlapping wavelength region of $8350-9200$\AA . The $9200-10000$\AA\ region is affected by telluric contamination, particularly in the Lick optical spectrum obtained on 2023 May 11 due to high-humidity conditions of that night, and was therefore excluded from the flux calibration. The magnitude offsets in the optical–NIR overlap region at different epochs were approximated using the corresponding changes in the $i$-band magnitudes. We caution, however, that between +80 and +101 day the line features such as Ca\,\two\ triplets may still undergo non-negligible evolution \citep{Friesen17, Tucker21}, which could introduce systematic biases in the absolute flux calibration. We adopt a linear flux-calibration relation and show the calibrated NIR spectra in physical units in Figure~\ref{fig:NIRcomparison}. On the other hand, owing to the large uncertainties in the overlapping region of the +333 day NIR spectrum below 10000\AA, we were unable to establish a reliable relative flux calibration with the +359 day optical spectrum. We therefore exclude this spectrum from subsequent analyses that rely on an accurate absolute flux scale. 

\begin{table}
\caption{Log of Spectroscopic Observations of SN~2023bee. }

\centering
 \begin{tabular}{c c c c} 
 \hline
 Obs Date & MJD & \tablenotemark{a}Phase & Telescope/Instrument \\ 
 (UT) &  & (Days) &\\ [0.3ex] 
 \hline\hline
2023-05-11 & 60075.2& 82.0 & Shane/Kast\\
2023-06-01 & 60096.98& 103.7 & Magellan/FIRE\\
2024-01-19 & 60329.12& 334.3 & Magellan/FIRE\\
2024-02-14 & 60354.4& 359.4 & Keck/LRIS\\

\hline
\end{tabular}
\tablenotetext{a}{Phases relative to B-band maximum on MJD 59992.58 in rest-frame.}
\label{tab:spec}
\end{table}
\section{Narrow H/He Features and non-degenerate companion}\label{sec:H/He}

\begin{figure*}
    \centering
    \includegraphics[width=0.99\linewidth]{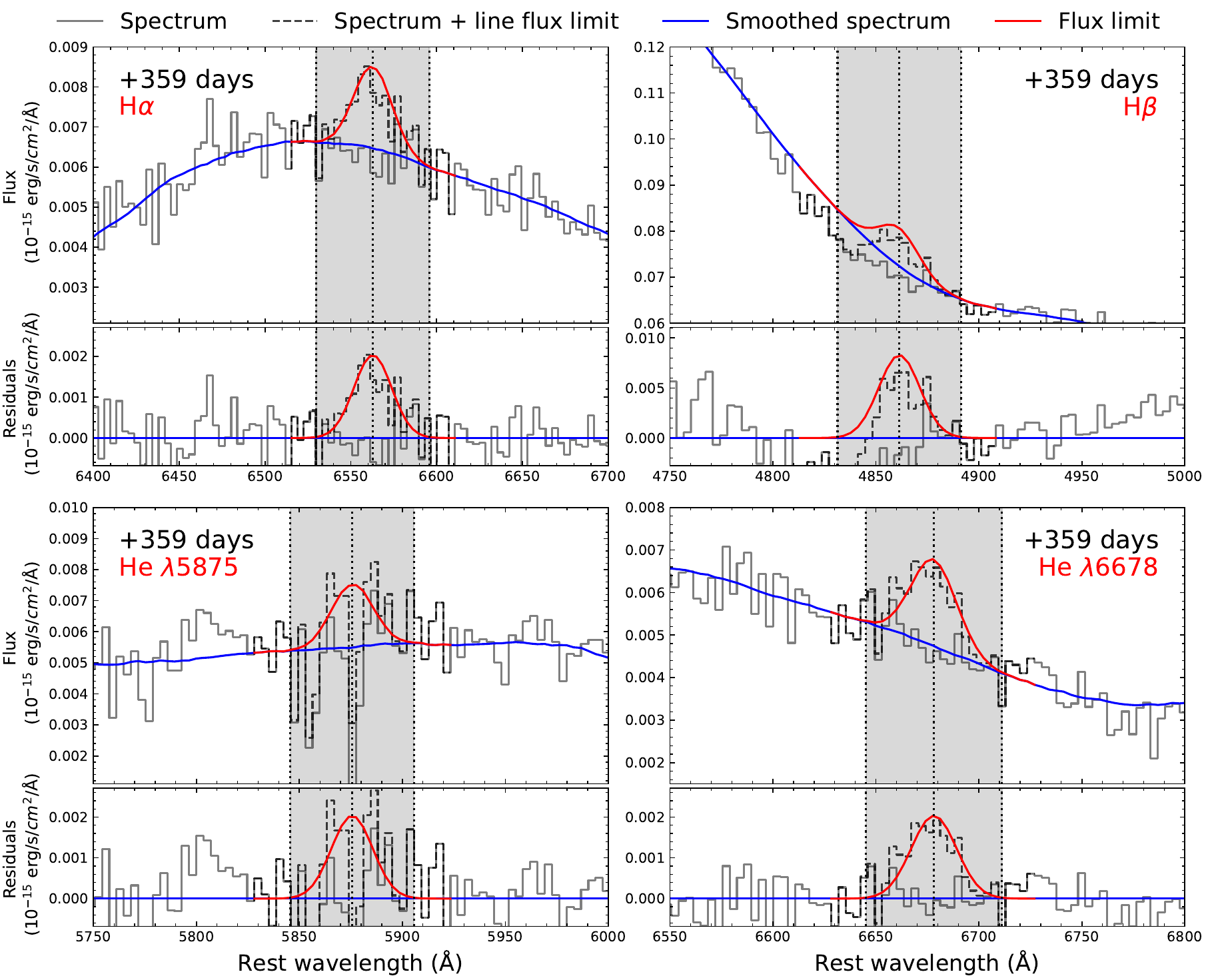}
    \caption{The LRIS optical spectrum, shown stepwise in black and smoothed in blue, zoomed in around the H$\alpha$, H$\beta$, He\,\one\ $\lambda5875$ and He\,\one\ $\lambda6678$ line regions. The spectra have been rebinned to the spectral resolution of the corresponding spectrograph. The red profiles show artificial emission features inserted at the rest wavelength of each transition, with a FWHM of $1500\ {\rm km,s^{-1}}$ and amplitudes corresponding to the $3\sigma$ detection threshold above the smoothed continuum. The corresponding appearance of each injected feature in the observed spectrum is plotted as the dashed black curves. The residuals with respect to the adopted continuum are also shown for each line. }
    \label{fig:Haline}
\end{figure*}

Figure~\ref{fig:Haline} shows zoom-ins for the $+359$ day LRIS spectrum for the H$\alpha$, H$\beta$, He\,\one\ $\lambda5875$ and He\,\one\ $\lambda6678$ lines. No prominent emission features are found at the expected wavelengths of relevant H and He features in optical and NIR in SN~2023bee.
Therefore, we follow the methodology of \cite{Dimitriadis2019} to place upper limits on H$\alpha$, H$\beta$, Paschen lines, and He\,\one\ emission and thereby constrain the amount of stripped material. The LRIS spectrum is rebinned to $3.5$\AA\ to approximate LRIS spectral resolution. We estimate the spectral uncertainties from the residual obtained by subtracting a smoothed continuum of the observed spectrum and use them to derive quantitative upper limits on undetected line emission at these wavelengths. To determine an appropriate continuum while minimizing the influence of any potentially present weak, narrow emission lines, we apply a second-order Savitzky--Golay filter with a broad window of $260$\AA.

The LRIS optical spectrum obtained on $+359$ day were observed in red and blue filters separately, separated at around $5700$\AA. We therefore estimate the residual noise independently within the blue and red wavelength ranges. Specifically, we calculate the standard deviation of the residuals over $4160$--$5560$\AA\ for the blue spectrum and $5700$--$9200$\AA\ for the red spectrum. These intervals encompass the H$\alpha$, H$\beta$, Pachen lines, and He\,\one\ $\lambda5875$ and He\,\one\ $\lambda6678$ regions, and the uncertainty levels are relatively uniform within each of the blue or red spectral regions. We therefore adopt the global uncertainties measured in red and blue arms rather than the local uncertainties around the individual line features. To avoid biasing the noise estimate from any weak emission at the expected line positions, we exclude $\pm100$\AA\ around each transition. We obtain a $3\sigma$ flux-density upper limit of $8.3\times10^{-18}$ erg s$^{-1}$ cm$^{-2}$ \AA$^{-1}$ for H$\beta$ in the blue filter. For the red filter, we find a corresponding $3\sigma$ upper limit of $2.0\times10^{-18}$ erg s$^{-1}$ cm$^{-2}$ \AA$^{-1}$ for H$\alpha$, He\,\one\,$\lambda5875$, and He\,\one\,$\lambda6678$. 

To visualize the limit, we add narrow Gaussian profiles on the optical spectrum at the rest wavelengths of H$\alpha$, H$\beta$, and He\,\one\ $\lambda5875$ and He\,\one\ $\lambda6678$ line regions. The widths of the Gaussian profiles are fixed at values corresponding to $1500\ \mathrm{km\,s^{-1}}$ at the wavelengths of the respective emission lines, equivalent to $\sim 33$\AA\ for H$\alpha$. The bottom panels of subplots in Figure~\ref{fig:Haline} show the comparison between the continuum-subtracted residuals and those artificial narrow Gaussian profiles representative of the corresponding flux limits. The comparison confirms that no significant narrow emission features are present at the wavelengths of interest. 
The resulting flux upper limits are $5.4, 1.6, 1.8,$ and $1.8\times10^{37}$ erg s$^{-1}$ for H$\beta$ He\,\one\,$\lambda5875$, H$\alpha$, and He\,\one\,$\lambda6678$. We caution, however, that the H$\beta$ limit is conservative. As shown in the upper right panel of Figure~\ref{fig:Haline}, the broad Fe features in the vicinity of the H$\beta$ region introduce additional residuals relative to the smoothed continuum. Consequently, the estimated uncertainty near H$\beta$ line is an overestimate of the intrinsic spectral noise. 

\cite{Botyanszki2018} derived the empirical relation between the line intensity and amount of stripped mass with an analytical fit to the multidimensional non-LTE radiative-transfer calculations basd on ejecta–companion interaction model in the Roche-lobe-filling scenario \citep{Boehner18}. With Equation 1 in \cite{Botyanszki2018},
\begin{equation}
    \log_{10}(L_{\text{H}\alpha})=-0.2M_1^2+0.17M_1+40,
\end{equation}
where $M_1=\log_{10}(M_{\text{st}}/M_\odot)$ and $[L_{\text{H}\alpha}]$ is in unit of $\text{erg s}^{-1}$, we estimate the upper limit of stripped H to be $M^\textrm{stripped}_{H} \lesssim 5.0\times10^{-4}M_\odot$. 
\cite{Botyanszki2018} also estimated that the same equation will overestimate the mass of stripped He by a factor of four in a He-only model. Therefore, we place an upper limit of the stripped He as $M^\textrm{stripped}_{He} \lesssim 1.2\times10^{-4}M_\odot$. 

\cite{Dessart20} pointed out that, if the stripped material is mixed within the ejecta when the spectra were taken, the optical depth will largely suppress H\,\one\ features, making H$\beta$ almost vanish before +300 days and decreasing H$\alpha$ intensity by a factor of $3-5$. Using the polynomial index fitted for three different simulation models at +300 days \citep[see Appendix A of][]{Dessart20}, we estimated an upper limit $M^\textrm{stripped}_{H} \lesssim 5.0\times10^{-3}M_\odot$ with the H$\alpha$ feature in these scenarios. A similar effect also influences the He\,\one\ features in optical, although, to our knowledge, no numerical models have yet quantified its magnitude in the case of He-rich ejecta.

\begin{figure*}
    \centering
    \includegraphics[width=0.99\linewidth]{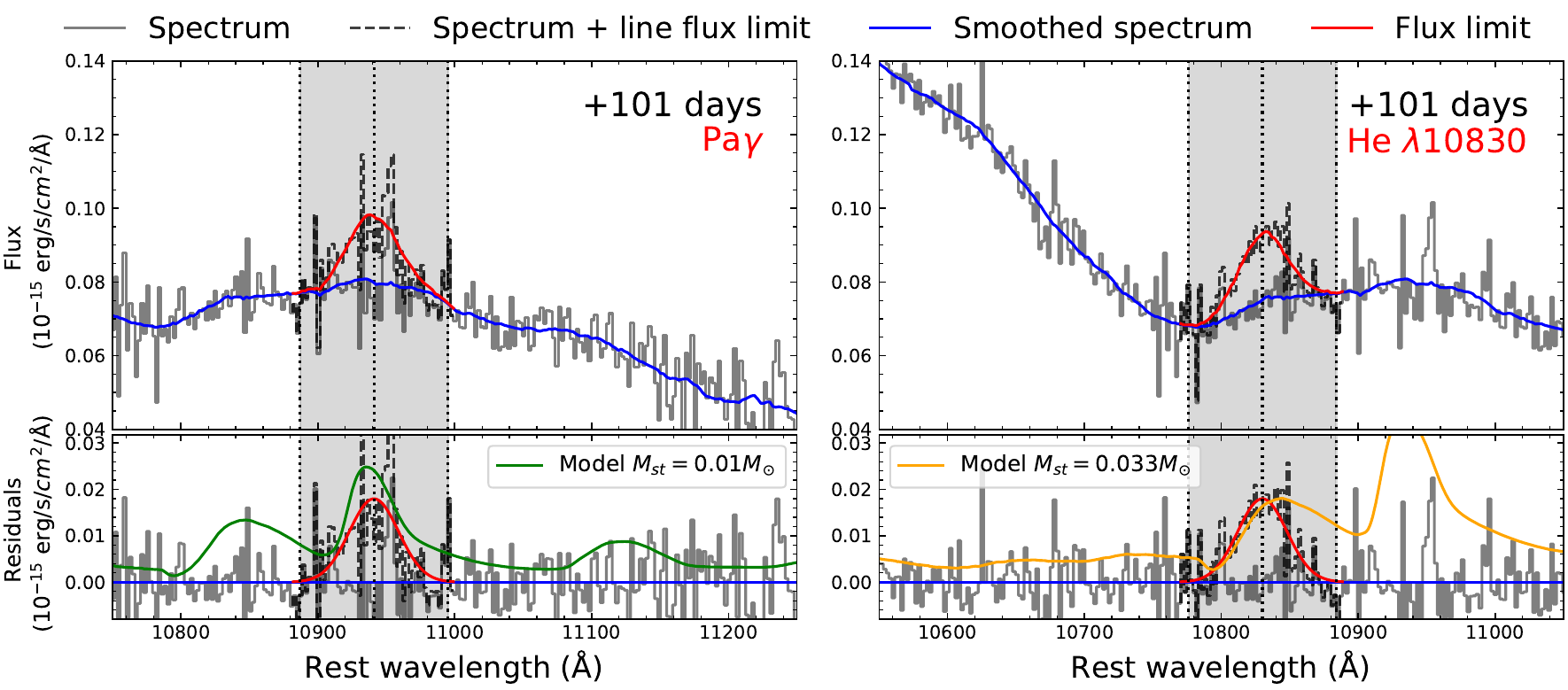}
    \caption{Same as Figure~\ref{fig:Haline}, but for the FIRE NIR spectrum taken at +101 day relative to the maximum, zooming in the Pa$\gamma$ and He\,\one\ $\lambda10830$ line regions. The bottom panels also show the expected line emission predicted by the \cite{Dessart20} models for $M_\textrm{st}\lesssim0.01M_\odot$ and $0.033M_\odot$, after subtracting the model with the minimal stripped mass, $M_\textrm{st}=1\times 10^{-5}M_\odot$, as the continuum baseline. }
    \label{fig:NIRline}
\end{figure*}

Additionally, \cite{Dessart20} simulated nebular phase NIR spectra of SNe~Ia in the H-rich companion scenario, including emission from stripped H/He-rich material, and predicted observable strong He\,\one\ 10830\AA\ and Pa$\gamma$. Figure~\ref{fig:NIRline} compare the FIRE NIR spectrum at +101 day with the corresponding +100 day models that most closely match the observational detection limits in the He\,\one\ 10830\AA\ and Pa$\gamma$ line regions. Following a similar approach, we derive an flux upper limit of $2.3\times10^{-18}$ erg s$^{-1}$ cm$^{-2}$ \AA$^{-1}$ for Pa$\gamma$ and He\,\one\ $\lambda10830$. In this comparison, the contribution from stripped materials as the excess flux relative to a baseline model with a minimal amount of stripped H-rich material, $M_\textrm{st} = 1\times10^{-5}M_\odot$. The +101 day NIR spectrum yields an upper limit of $M_\textrm{st}\lesssim0.01M_\odot$ for stripped H-rich material, which is less constraining than the limit derived from the +359 day optical spectrum. The constraint from the FIRE NIR spectrum at +333 days is even weaker given its low S/N and large uncertainties in absolute flux calibration. 


\section{Iron-group Line Diagnostics}\label{sec:7300A}

As discussed in Section~\ref{sec:intro}, the relative abundances of iron-group isotopes in the innermost ejecta are sensitive to the neutron excess and density attained during thermonuclear burning, and therefore provide important diagnostics of the explosion mechanism \citep{hoflich04, Blondin22}. Near-$M_{\rm Ch}$ models with steady accretion can reach central densities of $\sim10^9$ g cm$^{-3}$, substantially higher than the $\sim10^7$ g cm$^{-3}$ typical of sub-$M_{\rm Ch}$ models \citep{Rosswog09, Pakmor12, AlthausL+22}. The higher densities enhance electron capture and the production of stable neutron-rich iron-group isotopes, such as $^{54}$Fe and $^{58}$Ni, relative to radioactive $^{56}$Ni and its decay products \citep{hoflich04, Maguire18}. The [Ni,\two]/[Fe,\two] flux ratio has therefore been proposed as a probe of neutronization and, in turn, of the distinction between near-$M_{\rm Ch}$ and sub-$M_{\rm Ch}$ explosions \citep{flors20}.

This diagnostic becomes accessible at nebular phases, $\gtrsim200$ days after maximum light, when the ejecta are sufficiently optically thin to reveal emission from the iron-group-rich inner layers. By these epochs, the short-lived radioactive isotopes $^{56,57}$Ni have decayed to $^{56,57}$Co and to $^{56,57}$Fe as:
\[
\begin{aligned}
{}^{56}\mathrm{Ni}
&\xrightarrow{\,t_{1/2}=6.075~\mathrm{d}\,}
{}^{56}\mathrm{Co}
\xrightarrow{\,t_{1/2}=77.24~\mathrm{d}\,}
{}^{56}\mathrm{Fe},\\[4pt]
{}^{57}\mathrm{Ni}
&\xrightarrow{\,t_{1/2}=35.60~\mathrm{h}\,}
{}^{57}\mathrm{Co}
\xrightarrow{\,t_{1/2}=271.74~\mathrm{d}\,}
{}^{57}\mathrm{Fe}.
\end{aligned}
\]

At $\gtrsim300$ days, [Ni,\two] emission is expected to arise predominantly from stable neutron-rich Ni isotopes, particularly $^{58}$Ni, with a smaller contribution from $^{60}$Ni. In contrast, [Fe,\two] emission is dominated by the decay products of radioactive $^{56}$Ni, with additional contributions from stable Fe isotopes and other radioactive decay chains \citep{Graham17,Graham22,Maguire18,flors20,Siebert20,Seitenzahl13}. Because the production of stable neutron-rich iron-group isotopes depends strongly on the density and electron fraction of the burning material, near-$M_{\rm Ch}$ and sub-$M_{\rm Ch}$ models generally predict different Ni/Fe abundance ratios.
The [Fe,\two], [Ni,\two], and [Ca,\two] complex around $7300$~\AA\ can therefore constrain the stable Ni-to-Fe abundance ratio, with comparatively high Ni/Fe ratios generally associated with the enhanced neutronization expected in near-$M_{\rm Ch}$ explosions, whereas lower ratios are more readily reproduced by sub-$M_{\rm Ch}$ models \citep{flors20, Siebert23}. Complementary NIR transitions, particularly [Ni,\two] $\lambda1.939,\mu{\rm m}$ and [Fe,\two] lines such as $\lambda1.257,\mu{\rm m}$, provide additional constraints on the stable Ni abundance and the physical conditions of the iron-rich ejecta \citep{flors20,Blondin18,Blondin22,Blondin23}. 
In this work, we use the $\sim7200$~\AA\ complex together with NIR iron-group lines as an independent diagnostic of the explosion scenario. The inferred Ni/Fe ratio must be interpreted with caution though as the line intensities also depend on the ionization state and the mixing of different isotopes in the ejecta \citep{Blondin22,Blondin23}.


The line complex between $7000-7600$\AA\ is dominated by [Fe\,\two] ($7155, 7172, 7388, 7453$\AA) and [Ni\,\two] ($7378, 7412$\AA). [Co\,\three] ($6853, 6961, 7152$\AA) could also contribute to this region, particularly on its blue side, thereby affecting the inferred [Ni\,\two] and [Fe\,\two] line strengths at a non-negligible level \citep{flors20, Liu23}. Previous studies suggest that [Ca\,\two] ($7291, 7324$\AA) is another component that could contribute to this region \citep{Mazzali15, Fransson15,Botyanszki17}, though a number of empirical studies have put strong upper limits on its existence in other SNe~Ia \citep{Maguire18, Graham17}. 
In the following analysis, we assume that the observed emission is dominated by these transitions and lines arising from the same ion share a common velocity offset and line width. The rest wavelength and Einstein coefficient of those lines are taken from the NIST Atomic Spectra Database.\footnote{\url{https://www.nist.gov/pml/atomic-spectra-database}}

\begin{figure}
    \centering
    \includegraphics[width=0.99\linewidth]{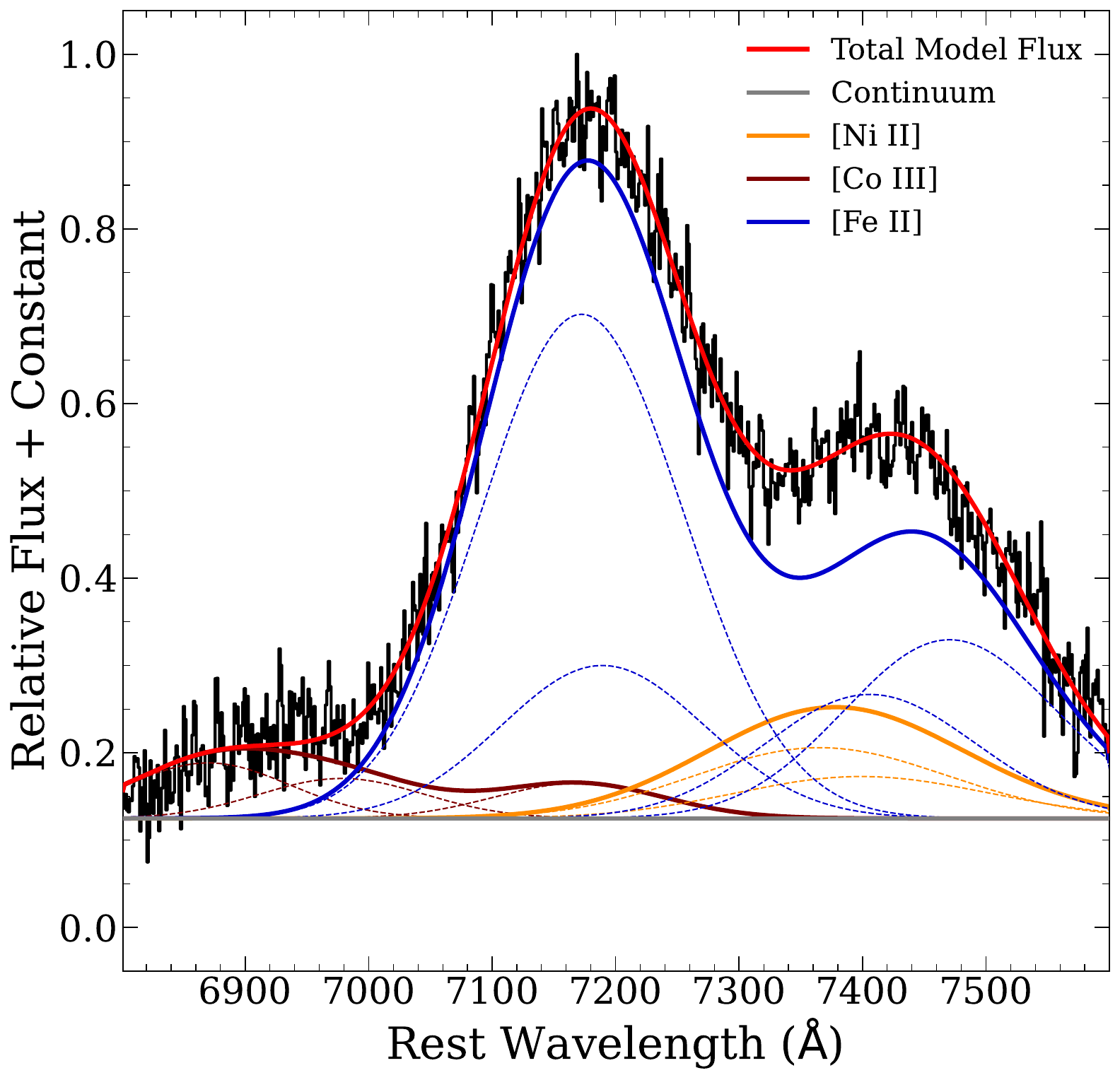}
    \caption{Best-fit for the $7300$\AA\ region with [Fe\,\two] (blue), [Ni\,\two] (orange) and [Co\,\three] (brown) with constant continuum (grey). Dashed lines are the individual emission lines, and solid lines represent the combination of certain ion. The flux is normalized to the $7180$\AA\ peak. The measured line ratio between [Ni\,\two] $\lambda7378$ and [Fe\,\two] $\lambda7155$ is used to constrain the the relative abundance of Ni\,\two\ and Fe\,\two.}
    \label{fig:7300fit}
\end{figure}

Figure~\ref{fig:7300fit} shows a zoom-in plot for the optical LRIS spectrum taken on +359 day in the the $\sim7200$~\AA\ complex region, normalized to the peak flux in this region. We fit this feature using several multi-component emission-line models: Fe+Ni, Fe+Co+Ni, and Fe+Co+Ni+Ca. For configurations including [Co\,\three], we extend the blue boundary of the fitting region to $\sim6800$~\AA\ in order to include the [Co\,\three] features at $\lambda6854$ and $\lambda6961$, while the fitting range of Fe+Ni model is restricted $7000-7600$\AA\ to exclude [Co\,\three]. Our fitting procedure broadly follows the approach of \cite{Siebert23}, with one important modification in the treatment of local continuum. In \cite{Siebert23}, the continuum was estimated by linearly interpolating between spectral regions on either side of the emission complex. This approach can bias the inferred line fluxes when the adopted continuum regions themselves contain weak line emission, particularly on the blue side of the complex where [Co\,\three] may contribute. We therefore adopt a constant continuum level across the fitting region and treat it as a free parameter.


The results for different fitting configurations are listed in Table~\ref{tab:Fefit}. 
We compare the fitting results using both the reduced $\chi^2$ and the Bayesian information criterion (BIC),. For Fe+Ni+Co and Fe+Ni+Co+Ca, we additionally report the reduced $\chi^2$ over both original fitting ranges and between $7000-7600$\AA\ to enable a direct comparison with Fe+Ni model. The BIC is used primarily to compare Fe+Ni+Co and Fe+Ni+Co+Ca with different number of parameters, and the BIC for Fe+Ni is provided only for reference as the fitting range differs.
In general, the Fe+Co+Ni model with a constant continuum shows the best performance while yielding consistent $v_\textrm{offset}$ and line width among the iron-group species, and is therefore our preferred configuration. 
This result indicates that [Co\,\three] makes a non-negligible contribution to the blended emission complex. When [Co\,\three] is omitted, its contribution near the blue side of the complex is partially absorbed into the fitted [Fe\,\two] component. This changes the inferred [Fe\,\two] contribution at longer wavelengths and, in turn, reduces the flux attributed to [Ni\,\two] $\lambda7378$. Consequently, models that neglect [Co,\three] yield systematically lower inferred Ni/Fe ratios.
Including [Ca\,\two] produces somewhat more similar velocity offsets for the [Ni\,\two] and [Fe\,\two], but does not yield improvement in either $\chi^2$ or the BIC. Moreover, the inferred [Ca\,\two] intensity remains below the $2\sigma$ level. We therefore find no statistically significant evidence for [Ca\,\two] emission in this region, consistent with previous studies \citep{Maguire18, Graham17}.
We also test a linear continuum for each line-identification configuration. In all cases, the inferred continuum slopes are consistent with zero within the uncertainties. Allowing a linear continuum does not significantly improve either $\chi^2$ or the BIC, nor does it produce statistically significant changes in the best-fit [Fe,\two] or [Ni,\two] parameters.


We estimate the Ni-to-Fe abundance ratio following the approach of \citet{Maguire18}. Because Ni\,\two\ and Fe\,\two\ have similar ionization potentials, their number-density ratio is expected to approximately trace the elemental Ni/Fe abundance ratio \citep{Fransson15}
Assuming optically thin emission, the ratio of the [Ni\,\two] $\lambda7378$ and [Fe\,\two] $\lambda7155$ line fluxes follows the equation: 
\begin{equation}
    \frac{L_{7378}}{L_{7155}} = 4.9 \frac{n_{\rm Ni II}}{n_{\rm Fe II}} \exp{\left(\frac{0.28~\mbox{eV}}{kT}\right)}\frac{dc_{\rm Ni II}}{dc_{\rm Fe II}}~.
\end{equation}
where ${L_{7378}}$ and ${L_{7155}}$ are the measured flux ratio of [Ni\,\two] 7378\AA\ and [Fe\,\two] 7155\AA, respectively. $dc_{\rm Ni II}/{dc_{\rm Fe II}}$ is the departure coefficients from LTE for the appropriate states of Ni\,\two\ and Fe\,\two. Following \cite{Maguire18}, we adopt a mean value $dc_{\rm Ni II}/{dc_{\rm Fe II}}=1.8\pm0.6$ estimated from the W7 model at +330d 
and allow the temperature to vary over $3000-8000$K in nebular phase \citep{Fransson15}. The range of the inferred $n_{\rm Ni II}/{n_{\rm Fe II}}$ for different line-identification models are listed in Table~\ref{tab:Fefit}.


NIR transitions of iron-group ions can provide additional constraints on the intensity, velocity, and width of optical features arising from the same ions. Using a sample of nebular-phase optical+NIR X-Shooter spectra of SNe~Ia, \cite{flors20} found that the flux ratio of the [Fe\,\two] $\lambda7155$ and $\lambda12570$ lines follows an approximately log-linear temporal evolution with time since explosion:
\begin{align}
    \textrm{log}\frac{L_{12570}}{L_{7155}} = &-(1.65\pm0.07)+(0.0043\pm0.0002)\,d^{-1} \notag\\
    & \times t_{exp}\textrm{[days]}~ 
\end{align}
with an intrinsic scatter of 0.06 dex. This trend may be explained by the natural cooling and dilution of the ejecta. Because of the relatively large uncertainties associated with the optical--NIR flux calibration and with the empirical relation itself, we do not impose this relation as a formal prior. Instead, we use it only as an independent consistency check on the inferred intensity and width of [Fe\,\two] $\lambda7155$ in the $7200$~\AA\ line complex. From the flux calibrated FIRE+LRIS spectrum, we measure a line intensity ratio of $F_{12570}/F_{7155} = 0.76\pm0.07$, in good agreement with the expected ratio between +350-380 days after explosion. 
In principle, the strength of Ni\,\two\ $\lambda7378$ can similarly be constrained using Ni\,\two\ $\lambda19390$ \citep{flors20, Dhawan18}. In our case, however, the relatively low S/N and the strong telluric absorption blueward of Ni\,\two\ $\lambda19390$ preclude a robust measurement of this feature.

Overall, the constant-continuum models are preferred, and the best-fit Fe+Ni+Co configuration yields a small $n_{\rm Ni II}/{n_{\rm Fe II}}=0.026\pm0.020$. Following the treatment in \cite{flors20}, we account for the continued conversion of radioactive Co into Fe and scale the inferred abundance ratio at the observed epoch to its asymptotic value at $t\rightarrow\infty$. We thereby obtain an upper limit on the asymptotic Ni-to-Fe mass ratio of $M_{\rm Ni}/{M_{\rm Fe}}\lesssim0.05$. Adding Ca\,\two] to the line model is disfavored by the BIC and does not significantly improve the quality of the fit. Even when Ca\,\two] is included, the inferred abundance remains consistent with $M_{\rm Ni}/M_{\rm Fe}\lesssim0.06$ at $t\rightarrow\infty$.  
However, this result should be interpreted with caution. As shown in the comparison between different fitting settings, the best-fit result is highly dependent on the estimate of the continuum and the S/N because this line complex is highly blended. 
Nonetheless, the fittings we have largely exhausted the potential configurations and range of ion ratio, and we can conservatively conclude that the upper limit of $\textrm{M}_{\textrm{Ni}}/\textrm{M}_{\textrm{Fe}}$ is $\sim 0.06$ at $t_\infty$. In both cases, the small Ni-to-Fe mass ratio is consistent with a sub-M$_\text{Ch}$ Double Detonation progenitor model \citep{Shen18, flors20, Kushnir_Waxman20}. The implications for the progenitor systems will be further discussed in Section~\ref{sec:discussion}.






\begin{table*}[]
    \centering
    \caption{Best fit parameters for different line configuration fits to the line profile around $7300$\AA. The continuum is is normalized to the $7180$\AA\ peak. The relative amplitude is also normalized to the $7180$\AA\ peak and then subtracted by continuum. Notice that the main uncertainty source for physical amplitude is the uncertainty in the flux calibration. }

\makebox[1.8\columnwidth][c]{%
\hspace*{-2cm}
\begin{tabular}{cccccccccc}
    \hline
      Model & fitting range & Lines & Relative Amplitude & $v_\textrm{offset}$  & Width & Continuum & $n_{\rm Ni II}/{n_{\rm Fe II}}$ & $\chi^2_\nu$ \tablenotemark{a} & BIC \tablenotemark{b} \\
     & [\AA] & & & [km s$^{-1}$] & [km s$^{-1}$] & & & &\\
      \hline      \hline
      Fe+Ni & [7000, 7600] & [Fe\,\two] & $0.61\pm0.01$& $875\pm55$ & $3490\pm100$ & $0.15\pm0.01$ & $0.0125\pm0.009$& 176& -500\\
       & & [Ni\,\two]  & $0.063\pm0.006$& $-50\pm470$ & $1880\pm420$  & & & &\\
      \hline
      Fe+Ni+Co & [6800, 7600] & [Fe\,\two]  & $0.60\pm0.01$ & $800\pm120$ & $3340\pm180$ & $0.15\pm0.03$& $0.026\pm0.02$ & 168 & -616\\
       & & [Ni\,\two]  &  $0.077\pm0.009$ & $-470\pm940$ & $3000\pm1200$ & & & (250)&\\
       & & [Co\,\three]  & $0.97\pm0.33$ & $2300\pm1500$ & $2350\pm660$  & & & &\\
      \hline
      Fe+Ni+Co+Ca & [6800, 7600] & [Fe\,\two]  & $0.59\pm0.01$ & $680\pm110$ & $3360\pm110$ & $0.117\pm0.024$ & $0.035\pm0.026$& 171 & -592\\
       & & [Ni\,\two]  &  $0.069\pm0.019$ & $600\pm280$ & $4460\pm830$ & & & (252) & \\
       & & [Co\,\three]  & $1.21\pm0.35$ & $980\pm530$ & $2820\pm500$   & & & & \\
       & & Ca\,\two]  & $0.030\pm0.019$ & $-150\pm590$ & $2800\pm1100$ &  & & & \\

       \hline
    \end{tabular}
    }
    \tablenotetext{a}{%
    \raggedright
    $\chi^2_\nu$ is evaluated over common fitting range $7000-7600$\AA. Values in parentheses are evaluated over $6800-7600$\AA.}
\tablenotetext{b}{%
\raggedright
BIC is calculated over the fitting range of each model. The BIC of the Fe+Ni model is therefore provided for reference only and not directly comparable to those of the other models.}

    \label{tab:Fefit}
\end{table*}

\section{Discussion}\label{sec:discussion}



Early-time observations of SN~2023bee revealed a pronounced flux excess within the first few days after explosion, with proposed origins including ejecta--companion interaction, circumstellar interaction, shallow or surface $^{56}$Ni, and a sub-$M_{\rm Ch}$ double-detonation explosion \citep{Hosseinzadeh2023,Wang23}. Both \citet{Hosseinzadeh2023} and \citet{Wang23} found that ejecta--companion interaction can provide a reasonable description of the early optical emission, with inferred binary separations of order $a\sim(1$--$3)\times10^{11}\,\mathrm{cm}$ under the assumption of a Roche-lobe-filling non-degenerate companion, corresponding approximately to a main-sequence companion rather than a red giant. However, \citet{Wang23} showed that the companion-interaction model does not simultaneously reproduce the full UV--optical evolution. They further explored alternative scenarios, finding that shallow $^{56}$Ni extending into the outer $\sim0.1$--$0.2\,M_\odot$ of the ejecta can reproduce the early optical rise, while a sub-$M_{\rm Ch}$ double-detonation model with a $\sim1.0\,M_\odot$ CO white dwarf and a thin $\sim0.01$--$0.02\,M_\odot$ He shell provides a comparably good description of the early light-curve morphology. Thus, although companion interaction remains a viable interpretation of the early excess, the early-time observations alone do not uniquely determine its physical origin and motivate independent constraints from the nebular phase.

In our late-time spectra, we find no significant narrow H or He emission associated with material stripped from a non-degenerate companion. Using the \citet{Botyanszki2018} prescription, we derive nominal upper limits of $M^{\rm stripped}_{\rm H}\lesssim5.0\times10^{-4}\,M_\odot$ and $M^{\rm stripped}_{\rm He}\lesssim1.2\times10^{-4}\,M_\odot$. Accounting for the suppression of optical H emission by ejecta optical-depth effects following \citet{Dessart20} weakens the more conservative hydrogen constraint to $M^{\rm stripped}_{\rm H}\lesssim5.0\times10^{-3}\,M_\odot$. These limits remain well below the stripped masses predicted by ejecta--companion interaction calculations. For example, \citet{Boehner18} found $\sim0.24$--$0.37\,M_\odot$ of stripped material for main-sequence companions with masses of $1.13$--$1.53\,M_\odot$ and binary separations of $2.13$--$6.58\,R_\odot$, partially overlapping the parameter space relevant to the companion-interaction interpretation of SN~2023bee. The observed limits therefore strongly disfavor such a classical single-degenerate configuration. Similar non-detections have been reported for several other SNe~Ia with early flux excesses, including SNe~2017cbv, 2018oh, 2019yvq, and 2021aefx \citep{Sand2018,Dimitriadis2019,Tucker19,Siebert20,Tucker21,Hosseinzadeh22}. In light of these repeated non-detections in SNe~Ia that otherwise show early-time flux excesses, companion interaction with a hydrogen- or helium-rich non-degenerate star is unlikely to be the dominant origin of early excesses in the population. More generally, systematic searches of nebular spectra have shown that narrow H/He emission is rarely detected in normal SNe~Ia \citep{Tucker20}, disfavoring the SD progenitor scenario as the common origin of normal SNe~Ia.

The nebular iron-group emission provides an independent constraint on the explosion conditions. At late phases, [Fe\,\two] emission is dominated by the decay products of radioactive $^{56}$Ni, whereas [Ni\,\two] primarily traces stable Ni synthesized during the explosion. The relative strength of these species therefore probes the degree of neutronization and, consequently, the central density of the pre-explosion progenitor. Modeling of the blended $\sim7300$\AA\ complex in SN~2023bee yields a weak [Ni\,\two] contribution, with $L_{7378}/L_{7155}\sim0.1$--$0.2$ and a conservative asymptotic Ni-to-Fe mass-ratio limit of $M_{\rm Ni}/M_{\rm Fe}\lesssim0.06$. This places SN~2023bee toward the low-Ni end of the distribution measured in nebular samples \citep{Maguire18,Graham22} and is more naturally reproduced by relatively low-density, sub-$M_{\rm Ch}$ burning than by standard high-density near-$M_{\rm Ch}$ delayed-detonation models \citep{flors20}. However, the inferred abundance ratio and physical interpretation remain subject to substantial systematic uncertainty. In particular, the method of \citet{flors20} does not account for variations in the chemical composition or density structure of the inner ejecta and assume that each ion has a free density. As \cite{Blondin22} pointed out, the intensity of [Ni\,\two] is also dependent on the ionization status and mixing of isotopes. For example, the lower density of [Ni\,\two] in sub-$M_{\rm Ch}$ model suppresses the recombination process of Ni ions, resulting less Ni$^+$ ions and thus lower [Ni\,\two] intensity compared to $M_{\rm Ch}$ explosion. In addition, the strong blending in the 7300\AA\ region makes the result sensitive to the continuum placement, the adopted [Co\,\three] and [Ca\,\two] contributions, and the relatively modest S/N of the spectrum. We therefore regard the low Ni/Fe ratio primarily as evidence favoring lower-density burning rather than as a unique identification of a particular explosion mechanism.

\section{Conclusion}\label{sec:conclusion}

We present late-time optical and NIR spectroscopy of SN~2023bee to investigate the progenitor and explosion scenario of this SN~Ia with a prominent early flux excess. We find no significant narrow H or He emission associated with material stripped from a non-degenerate companion, placing upper limits of $M^{\rm stripped}_{\rm H}\lesssim5.0\times10^{-3}\,M_\odot$ and $M^{\rm stripped}_{\rm He}\lesssim1.2\times10^{-4}\,M_\odot$. These limits are difficult to reconcile with the non-degenerate companion configuration required by a simple ejecta–companion interpretation of the early excess. Modeling of the nebular $\sim7300$\AA\ iron-group complex further yields a weak [Ni,\two] contribution and a conservative Ni-to-Fe mass-ratio limit of $M_{\rm Ni}/M_{\rm Fe}\lesssim0.06$, favoring a lower-density, sub-$M_{\rm Ch}$-like explosion, although the inference remains subject to systematic uncertainties in the modeling. Taken together, these late-time constraints provide a complementary view of the progenitor and explosion physics of SN~2023bee beyond that available from the early light curve alone. While no single diagnostic uniquely determines the progenitor scenario, the combined absence of stripped companion material and weak stable-Ni signature is consistent with a sub-$M_{\rm Ch}$ explosion and motivates consideration of a double-detonation origin for the early excess. 


More broadly, nebular observations increasingly point toward a heterogeneous SN~Ia population: stringent limits on stripped material place strong constraints on classical single-degenerate systems, while the diversity of nebular iron-group abundances indicates a range of central densities and explosion conditions. Recent nebular spectroscopy with \textit{JWST} has revealed strong stable-Ni emission and signatures of high-density burning in several normal SNe~Ia, including SNe~2021aefx, SN~2022aaiq, and SN~2024gy, broadly consistent with near-$M_{\rm Ch}$ explosions \citep{Kwok23,DerKacy23,Ashall24,Kwok26}. In contrast, different analyses of SN~2022xkq have favored a near-$M_{\rm Ch}$ delayed-detonation explosion \citep{DerKacy24} and a sub-$M_{\rm Ch}$ explosion \citep{Kwok26}, respectively, which highlights the sensitivity of progenitor inferences from nebular spectra to the reduction method and spectroscopic modeling treatments. Such diversity suggests that multiple progenitor and explosion pathways likely contribute to the observed SN~Ia population. 

Establishing how these late-time properties relate to early light-curve behavior will require larger samples combining high-cadence observations during the first days after explosion with high-S/N nebular optical, NIR, and MIR spectroscopy. In particular, infrared observations provide access to less blended iron-group transitions and therefore cleaner constraints on stable isotopes and the physical conditions of the inner ejecta. Together with improved multidimensional non-LTE radiative-transfer calculations and atomic data, these observations will enable early-time signatures to be connected more directly with the composition, density, and geometry of the ejecta on an event-by-event and ultimately population-wide basis.

\section*{Acknowledgments}


This work was partially supported by NASA ADAP grant 80NSSC22K0494.

Q.W.\ is supported by the Sagol Weizmann-MIT Bridge Program.
M.R.S.\ is supported by the STScI Postdoctoral Fellowship. 
P.C.\ acknowledges support from the Zhejiang Provincial Top-Level Research Support Program.
Y.Z.\ acknowledges support from the MAOF grant (no.\ 12641898) and visitor support from the Observatories of the Carnegie Institution for Science, Pasadena, CA, where part of this work was completed.
K.W.D.\ is supported in part by a University of California President's Pre-Professoriate Fellowship (PPPF).
The UCSC team is supported in part by NASA grants 80NSSC23K0301 and 80NSSC24K1411; and a fellowship from the David and Lucile Packard Foundation to R.J.F.
A.G-Y.’s research is supported by ISF, IMOS and BSF grants, as well as the André Deloro Institute for Space and Optics Research, the Center for Experimental Physics, a WIS-MIT Sagol grant, the Norman E.\ Alexander Family M Foundation ULTRASAT Data Center Fund, and Magnus Konow in honor of his mother Olga Konow Rappaport;  A.G-Y.\ is the incumbent of the The Arlyn Imberman Professorial Chair.
W.B.H.\ acknowledges support from the National Science Foundation Graduate Research Fellowship Program under Grant No.\ 2236415.
C.D.K.\ gratefully acknowledges support from the NSF through AST-2432037, the HST Guest Observer Program through HST-SNAP-17070 and HST-GO-17706, and from JWST Archival Research through JWST-AR-6241 and JWST-AR-5441.
C.L.\ is supported by DoE award \#\,DE-SC0025599. 
G.N.\ is funded by NSF CAREER grant AST-2239364, supported in-part by funding from Charles Simonyi. G.N.\ also gratefully acknowledges NSF support from OAC-2311355, AST-2432428, as well as AST-2421845 and funding from the Simons Foundation for the NSF-Simons SkAI Institute. GN is also supported by the DOE through the Department of Physics at the University of Illinois, Urbana-Champaign.
A.S.\ is is supported by a VILLUM FONDEN Villum Experiment grant (VIL69896) and a research grant (VIL54489) from VILLUM FONDEN.

The Young Supernova Experiment (YSE) and its research infrastructure is supported by the European Research Council under the European Union's Horizon 2020 research and innovation programme (ERC Grant Agreement 101002652, PI K.\ Mandel), the Heising-Simons Foundation (2018-0913, PI R.\ Foley; 2018-0911, PI R.\ Margutti), NASA (NNG17PX03C, PI R.\ Foley), NSF (AST-1720756, AST-1815935, PI R.\ Foley; AST-1909796, AST-1944985, PI R.\ Margutti), the David \& Lucille Packard Foundation (PI R.\ Foley), VILLUM FONDEN (project 16599, PI J.\ Hjorth), and the Center for AstroPhysical Surveys (CAPS) at the National Center for Supercomputing Applications (NCSA) and the University of Illinois Urbana-Champaign.

Pan-STARRS is a project of the Institute for Astronomy of the University of Hawaii, and is supported by the NASA SSO Near Earth Observation Program under grants 80NSSC18K0971, NNX14AM74G, NNX12AR65G, NNX13AQ47G, NNX08AR22G, 80NSSC21K1572, and by the State of Hawaii.  The Pan-STARRS1 Surveys (PS1) and the PS1 public science archive have been made possible through contributions by the Institute for Astronomy, the University of Hawaii, the Pan-STARRS Project Office, the Max-Planck Society and its participating institutes, the Max Planck Institute for Astronomy, Heidelberg and the Max Planck Institute for Extraterrestrial Physics, Garching, The Johns Hopkins University, Durham University, the University of Edinburgh, the Queen's University Belfast, the Harvard-Smithsonian Center for Astrophysics, the Las Cumbres Observatory Global Telescope Network Incorporated, the National Central University of Taiwan, STScI, NASA under grant NNX08AR22G issued through the Planetary Science Division of the NASA Science Mission Directorate, NSF grant AST-1238877, the University of Maryland, Eotvos Lorand University (ELTE), the Los Alamos National Laboratory, and the Gordon and Betty Moore Foundation.

Some of the data presented herein were obtained at Keck Observatory, which is a private 501(c)3 non-profit organization operated as a scientific partnership among the California Institute of Technology, the University of California, and the National Aeronautics and Space Administration. The Observatory was made possible by the generous financial support of the W.\ M.\ Keck Foundation.
The authors wish to recognize and acknowledge the very significant cultural role and reverence that the summit of Maunakea has always had within the indigenous Hawaiian community.  We are most fortunate to have the opportunity to conduct observations from this mountain.

A major upgrade of the Kast spectrograph on the Shane 3~m telescope at Lick Observatory was made possible through generous gifts from the Heising-Simons Foundation as well as William and Marina Kast. Research at Lick Observatory is partially supported by a generous gift from Google.

This paper includes data gathered with the 6.5 meter Magellan Telescopes located at Las Campanas Observatory, Chile.

This work has made use of data from the Asteroid Terrestrial-impact Last Alert System (ATLAS) project. The Asteroid Terrestrial-impact Last Alert System (ATLAS) project is primarily funded to search for near earth asteroids through NASA grants NN12AR55G, 80NSSC18K0284, and 80NSSC18K1575; byproducts of the NEO search include images and catalogs from the survey area. This work was partially funded by Kepler/K2 grant J1944/80NSSC19K0112 and HST GO-15889, and STFC grants ST/T000198/1 and ST/S006109/1. The ATLAS science products have been made possible through the contributions of the University of Hawaii Institute for Astronomy, the Queen’s University Belfast, the Space Telescope Science Institute, the South African Astronomical Observatory, and The Millennium Institute of Astrophysics (MAS), Chile.

Based on observations obtained with the Samuel Oschin Telescope 48-inch and the 60-inch Telescope at the Palomar Observatory as part of the Zwicky Transient Facility project. ZTF is supported by the National Science Foundation under Grant Nos. AST-1440341, AST-2034437, and a collaboration including Caltech, IPAC, the Weizmann Institute for Science, the Oskar Klein Center at Stockholm University, the University of Maryland, the University of Washington, Deutsches Elektronen-Synchrotron and Humboldt University, the TANGO Consortium of Taiwan, the University of Wisconsin at Milwaukee, Trinity College Dublin, Lawrence Livermore National Laboratories, and IN2P3, France. Operations are conducted by COO, IPAC, and UW.

YSE-PZ was developed by the UC Santa Cruz Transients Team, supported in part by NASA grants NNG17PX03C, 80NSSC19K1386, and 80NSSC20K0953; NSF grants AST--1518052, AST--1815935, and AST--1911206; the Gordon \& Betty Moore Foundation; the Heising-Simons Foundation; a fellowship from the David and Lucile Packard Foundation to R.\ J.\ Foley; Gordon and Betty Moore Foundation postdoctoral fellowships and a NASA Einstein fellowship, as administered through the NASA Hubble Fellowship program and grant HST-HF2-51462.001, to D.\ O.\ Jones; and a National Science Foundation Graduate Research Fellowship, administered through grant No.\ DGE-1339067, to D.\ A.\ Coulter.


%

\vspace{5mm}
\facilities{ATLAS, Keck:I, Magellan:Baade, Nickel, PO:1.2m, PS1, Shane, Swope}


\software{astropy \citep{2013A&A...558A..33A,2018AJ....156..123A},
YSE-PZ \citep{Jones2021,Coulter2022, Coulter2023},
Matplotlib \citep{matplotlib}, 
SciPy \citep[][]{2020SciPy-NMeth}, 
NumPy \citep{2020NumPy-Array}, 
pyphot \citep[][]{fouesneaumorgan_2022},
emcee \citep{emcee}}





\bibliography{sample631}{}
\bibliographystyle{aasjournal}



\end{document}